\documentclass[manuscript]{aastex}
\usepackage{graphicx}

\slugcomment{}

\shorttitle{Molecular Gas in Void Galaxies}
\shortauthors{Das et al.}

\begin{document}


\title{Detection of Molecular Gas in Void Galaxies~:~Implications for Star Formation in Isolated Environments}


\author{M.~Das}
\affil{Indian Institute of Astrophysics, Bangalore, India}
\email{mousumi@iiap.res.in}

\author{T.~Saito}
\affil{Department of Astronomy, Graduate school of Science, The University of
Tokyo, 7-3-1 Hongo, Bunkyo-ku, Tokyo 133-0033, Japan}

\author{D.~Iono}
\affil{Chile Observatory, NAOJ, Japan}

\author{M.~Honey}
\affil{Indian Institute of Astrophysics, Bangalore, India}

\author{S.~Ramya}
\affil{Shanghai Astronomical Observatory, Shanghai, China}




\begin{abstract}
We present the detection of molecular gas from galaxies located in nearby voids using the CO(1--0) 
line emission as a tracer. The observations were done using the 45m single dish radio telescope 
of the Nobeyama Radio Observatory (NRO). Void galaxies lie in the most underdense parts of our universe
and  a significant fraction of them are gas rich, late type spiral galaxies. Although isolated, they 
have ongoing star formation but appear to be slowly evolving compared to galaxies in 
denser environments. Not much is known about their star formation properties or cold gas content. In 
this study we searched for molecular gas in five void galaxies. The galaxies 
were selected based on their relatively high IRAS fluxes or H$\alpha$ line 
luminosities, both of which signify ongoing star formation. All five galaxies appear to be isolated and
two lie within the Bootes void. We detected CO(1--0) emission from four of the five galaxies in our sample
and the molecular
gas masses lie between $10^{8} - 10^{9}~M_{\odot}$. We did followup H$\alpha$ imaging observations of three 
detected galaxies using the Himalayan Chandra Telescope and determined their star formation rates (SFR)
from their H$\alpha$ fluxes. The SFR varies from 0.2 - 1~M$_{\odot}~yr^{-1}$; which is similar to that observed
in local galaxies. Our study indicates that although void galaxies reside in underdense 
regions, their disks contain molecular gas and have star formation rates similar to galaxies
in denser environments. We discuss the implications of our results.  

\end{abstract}





\section{Introduction}

In the large scale structure of our universe, galaxies cluster along sheets, walls 
and filaments leaving large empty regions called voids in between 
\citep{kirshner.etal.1981,geller.huchra.1989,hoyle.vogeley.2004,foster.nelson.2009,sutter.etal.2012}. They 
represent the most under-dense parts of our universe and can vary in size but typical values are 10 to 20~Mpc. For
example the nearby Local Void is $\sim$23~Mpc across \citep{tully.etal.2008} whereas the Bootes Void, which is the largest 
known void, is 60~Mpc across \citep{kirshner.etal.1987}. 
In the past two decades, large optical surveys have revealed that voids contain a small but significant population of galaxies 
\citep{grogin.geller.2000,rojas.etal.2004,weygaert.2011}. The smaller voids 
are mainly populated by small, gas rich, low surface brightness (LSB) dwarfs \citep{chengalur.etal.2015} and irregular 
galaxies \citep{karachentseva.etal.1999} but the larger voids also support a population of relatively bright galaxies
that are late type, gas rich systems and often blue in color \citep{kreckel.etal.2011,ricciardelli.etal.2014,kreckel.etal.2015}. 
This is in contradiction to  
cold dark matter (CDM) models of structure formation, where voids are predicted to be mainly populated by low luminosity 
galaxies \citep{peebles.2001}. The presence of these blue void galaxies also indicates
that there is ongoing star formation in voids. Even some early type elliptical (E) and lenticular (S0) void galaxies 
(that are mainly composed of old stars) are found to be relatively blue 
in color \citep{wegner.grogin.2008}. However, although a signicant fraction of void galaxies do show star formation, they 
are in general evolving at a slower rate than galaxies in denser environments. This is evident from the color magnitude 
plot of these galaxies, in which the majority lie in the blue cloud region and not in the red region which is expected from
an evolved galaxy population \citep{kreckel.etal.2012}. 

One of the main attractions of studying void galaxies is that they provide us an opportunity to study star formation in 
isolated environments. It is not clear what drives star formation activity in such underdense environments. In denser 
parts of the universe, galaxy interactions and mergers play an important role 
in triggering star formation in gas rich galaxies. The interactions perturb the axisymmetric structure of the galaxy disks 
leading to enhanced cloud collisions, star formation and gas infall to the nuclear regions. The enhanced nuclear gas surface 
densities can lead to nuclear star formation, AGN activity and outflows. This cycle of activity ultimately results in bulge 
growth and the overall evolution of galaxies from small star forming blue systems to the more evolved red galaxies.  
In voids, galaxy interactions (such as the galaxy pair CG~693-692) are rare. Some void galaxies have low luminosity 
companions such as LSB dwarfs and dwarf spheroidals that are not easy to spot in optical images but nevertheless perturb 
galaxy disks and trigger star formation. Such interactions are difficult to detect in optical surveys. One way of detecting 
them is through HI surveys (e.g. Szomoru et al. 1997) where they may appear to have disturbed HI morphologies.
Although some void galaxies do show such features in their HI contours (e.g. BHI~1514+5155), interactions with visible or 
under-luminous companions are not high enough to explain the whole picture of star formation in voids and thus interactions 
may not be the main driver of star formation activity in these underdense environments \citep{grogin.geller.2000}. 
Another possible driver of star formation activity
in voids could be the slow accretion of cold gas by void galaxies from the inter-galactic medium (IGM) \citep{kreckel.etal.2012}. 
Gas accretion can enhance the gas surface densities in disks. This can result in the formation of local disk instabilities 
which lead to disk star formation \citep{dekel.birnboim.2006,keres.etal.2005}. Recent observations of HI gas filaments connecting 
galaxies in voids suggests that this process may be important for galaxy pairs or small groups 
\citep{beygu.etal.2013,kreckel.etal.2012,stanonik.etal.2009}. 

Void galaxies are also one of the only probes by which we can investigate the void substructure  - does it exist and how is it 
traced by galaxies? In $\Lambda$CDM theories of large scale structure (LSS) formation, voids evolve with time into larger, 
emptier volumes. In the process mass flows from voids into walls and filaments 
\citep{zeldovich.1970,icke.1984,bond.etal.1996,aragon-calvo.szalay.2013}. Isolated or small groups of 
galaxies are left behind within the voids, formed by the compression of filaments or walls as the voids merge 
\citep{sahni.etal.1994,sheth.weygaert.2004,weygaert.2010}. This void substructure, which in simulations appears as tenuous 
filaments and clusters of dark matter halos, can be traced by the distribution of void galaxies 
\citep{hahn.etal.2007,aragon-calvo.etal.2007,cautun.etal.2014}. 
Recent deep observations have shown that this void substructure exists \citep{alpaslan.etal.2014,kreckel.etal.2012}.  
But only really deep surveys will reveal the inner structure of voids and how it connects the galaxies residing 
within the voids \citep{penny.etal.2015}. As shown in simulations there will also be gas flowing along these filaments and accreting 
onto the void galaxies. This can lead to star formation in the galaxy disks. 
 
As a first step towards studying star formation and galaxy evolution in voids, we present a search for molecular gas in nearby 
void galaxies using the Nobeyama Radio telescope (NRO) using the CO(1--0) emission line as a tracer. Single dish observations will 
give an estimate of the molecular gas masses and the emission line profiles can reveal properties of the gas distribution, such as whether  
the gas is centrally concentrated or more extended in a rotating disk. There are only two earlier studies that have detected CO(1--0) in
voids; we use those results to enhance our sample. We followed up our CO(1--0) detections
with H$\alpha$ observations of three detected galaxies and derived their star formation rates.
In the following sections we describe our sample selection, the galaxy properties, observations and results.
For all distances we have used $H_{0}~=~73~km~s^{-1}~Mpc^{-1}$ and $\Omega~=~0.27$ \citep{komatsu.etal.2009}.

\section{Sample Selection}

We selected an initial sample of 12 galaxies from earlier studies of void  galaxies 
\citep{cruzen.etal.2002,weistrop.etal.1995,szomoru.etal.1996} and the Void Galaxy Survey (VGS) which used SDSS images to select
a sample of sixty void galaxies and study their HI/radio properties using the WSRT telescope \citep{kreckel.etal.2012}.
Our shortlisted galaxies had the following properties~: \\
(i)~Significant HI gas masses and stellar 
masses greater than $10^{9}~M_{\odot}$. This is to ensure avoiding dwarf galaxies in our sample since  
previous studies of low surface brightness (LSB) dwarfs indicate that they are unlikely to have 
molecular gas \citep{das.etal.2006}.\\
(ii)~Signatures of ongoing star formation. We checked the SDSS spectra of the sample galaxies to confirm the presence of 
emission lines that are  typical of star formation (e.g. H$\alpha$, H$\beta$ and [OIII]).\\
(iii)~For galaxies with IRAS fluxes, we selected only those that had $S_{100}~>~1.0~Jy$.\\
(iv)~None of the galaxies had been previously studied in CO(1--0) emission. 

However, due to unfavourable weather conditions we were able to observe only five galaxies from this sample. They are listed
in Table~1 and are briefly described below.\\
{\bf SBS1325+597 (IRASF~13254+5945, VGS~34)~:~}This is a gas rich galaxy with a size of $D_{25}\sim11.7~kpc$ (Table~1) 
that has been studied as part of the Void Galaxy Survey (here after VGS) \citep{kreckel.etal.2011}. The SDSS image of the galaxy 
shows a compact, red nucleus and a single, faint spiral arm. The SDSS optical spectrum has few emission lines 
suggesting only moderate star formation and shows no signatures of AGN activity. The HI is extended well beyond the optical radius 
and the outer isophotes are disturbed, which suggests that the galaxy maybe interacting with a distant companion galaxy.\\
{\bf SDSS~143052.33+551440.0 (SDSS1430+5514, VGS~44)~:~}This is also a moderate size disk galaxy with a $r$ band disk radius 
of $\sim3.6~kpc$ and an HI radius $<~7~kpc$ \citep{kreckel.etal.2012}. There is no strong bulge or extended disk but the SDSS image 
reveals a blue nucleus that shows strong H$\alpha$ emission in the optical spectrum, indicative of nuclear star formation.\\
{\bf SDSS~153821.22+331105.1 (SDSS1538+3311, VGS~57)~:~} This is the only galaxy in our sample that appears to have a bar.
It has an extended disk of radius $17.9^{\prime\prime}$ or $\sim8.1~kpc$ in the SDSS $r$ band image and an HI radius $<~9~kpc$ 
\citep{kreckel.etal.2012}. Though the galaxy is relatively blue and the SDSS nuclear spectrum shows strong H$\alpha$ emission, 
other signatures of strong star formation such as [OI] emission which is associated with shocks, are lacking. The ongoing star formation 
is located mainly along the bar.\\
{\bf CG~598 (IRAS~F14575+4228)~:~}This is a distant but large galaxy in the Bootes void that has disk radius of 
$22.4^{\prime\prime}$ or $\sim26.2~kpc$ in the $k_s$ band and shows signs of strong star formation in its optical spectrum 
(e.g. H$\alpha$, H$\beta$ and [OII] emission) \citep{cruzen.etal.2002}. It is gas rich and the HI gas disk is extended well beyond the optical 
disk Szomoru et al. (1996). In the SDSS $g$ band image, the galaxy appears to be accreting a smaller galaxy. The interaction 
probably triggered the burst of star formation.\\
{\bf SBS~1428+529 (IRAS~F14288+5255)~:~}This is also a distant galaxy located in the Bootes void. Like CG~598 it is large in size
(radius $\sim27^{\prime\prime}$ or $\sim24.6~kpc$) but appears to be isolated. It has a bright bulge, distinct spiral arms and 
a strong bar. It is the only galaxy in our sample that shows AGN activity;  it hosts a Seyfert~2 type nucleus \citep{cruzen.etal.2002}. 
 
\section{CO Observations and Data Reduction}

The $^{12}CO (J=1-0)$ single dish, emission line observations were done using the 45~m Nobeyama Radio Telescope
during the period 14~-~25 April, 2013. At the CO rest frequency of 115.271204~GHz, the half-power beam width (HPBW) is
15$^{\prime\prime}$ and the main beam efficieny is 30\%. We used the one beam (TZ1), dual polarization, sideband 
separating receiver (TZ). The signal was digitized to 3 bits before being transferred to the digital
FX-type spectrometer SAM45, that has a bandwidth of 4~GHz \citep{nakajima.etal.2008}.
Typical system temperatures were 160 - 260~K. The chopper-wheel method was used for temperature calibration with a swtiching 
cycle of 10~s. The resultant output had 4096 channels and a frequency resolution of 488~kHz.

The on source time for the first four galaxies varied between 1 to 1.5 hours. Due to poor weather conditions the fifth galaxy 
SBS~1428+529 was observed for only 25 minutes (Table~2). The pointing accuracy was about $\sim2^{\prime\prime}$ to $\sim4^{\prime\prime}$. 
The data was analysed using the NRO calibration tool NEWSTAR. Data with a wind velocity greater than $5~kms^{-1}$ and data points 
with winding baselines were flagged out. The antenna temperature ($T_A$) was converted to the main beam temperature ($T_{mb}$) using a main beam 
efficiency of $\eta_{b}$=0.3 where $T_{A}~=~T_{mb}/\eta_{b}$. The spectra were converted from Kelvin to Jansky using the conversion 
factor of 2.4~Jy~K$^{-1}$.  

\section{H$\alpha$ Optical Observations and Data Reduction}

We conducted H$\alpha$ observations of three galaxies in our sample, SBS~1325+597, SDSS~143052.33+551440.0 and SDSS~J153821.22+331105.1.
The remaining two galaxies were not observed because we did not have suitable filters.
The H$\alpha$ observations were done using the Himalayan Faint Object Spectrograph 
Camera (HFOSC) which is mounted on the 2m Himalayan Chandra Telescope (HCT) installed at the Indian Astronomical 
Observatory (IAO), Hanle, India. HFOSC has a 2K $\times$ 4K SITe CCD chip with a plate scale of 0.296 arcsec pixel$^{-1}$ . The 
central 2K $\times$ 2K covers a field of view of 10 $\times$ 10 arcmin$^{2}$. The observations were obtained 
on 2014 April 11 \& 25. For the galaxies SBS~1325+597 and SDSS 143052.33+551440.0 we used the H$\alpha$ broad filter of band width 
$\sim$~500~$\AA$ to obtain the H$\alpha$ images. But SDSS J153821.22+331105.1 is at a higher redshift of $z=0.023$ and the H$\alpha$ line 
is shifted to 6714~$\AA$. Hence for this galaxy we used the narrow-band [SII] filter of band width $\sim$~100~$\AA$ centered around the 
wavelength 6724~$\AA$. 

To obtain the continuum subtracted H$\alpha$ images we followed the procedure as outlined below. The galaxy images 
were obtained in both the broad-band $r$ filter and narrow-band filters centered around the H$\alpha$ 
line. The bias frames and twilight flats were used for preprocessing of the images. The data reduction was done 
using the standard packages available in IRAF\footnote{Image Reduction \& Analysis Facility Software distributed 
by National Optical Astronomy Observatories, which are operated by the Association of Universities for Research in 
Astronomy, Inc., under co-operative agreement with the National Science Foundation}. The frames were bias subtracted 
and flat field corrected using the master bias and master flat frames. The cosmic ray hits were removed using the 
task COSMICRAYS in IRAF. The images of the two different filters were aligned geometrically using the tasks GEOMAP \& GEOTRAN. 
The Point Spread Function (PSF) of both the broad and narrow-band frames were matched. The scale factor between the 
broad-band $r$ frame and the H$\alpha$ frame was determined using the field stars. The continuum subtracted 
H$\alpha$ images were obtained by subtracting the PSF matched scaled continuum $r$ band images from the 
narrow-band images as described in \citep{waller.1990}.  Flux calibration was done using the 
spectrophotometric standard star HZ44 \citep{oke.1990}. Flux calibrated continuum subtracted H$\alpha$ images are shown 
in Figures 3, 4 and 5.

\begin{figure}
\epsscale{1.2}
\plottwo{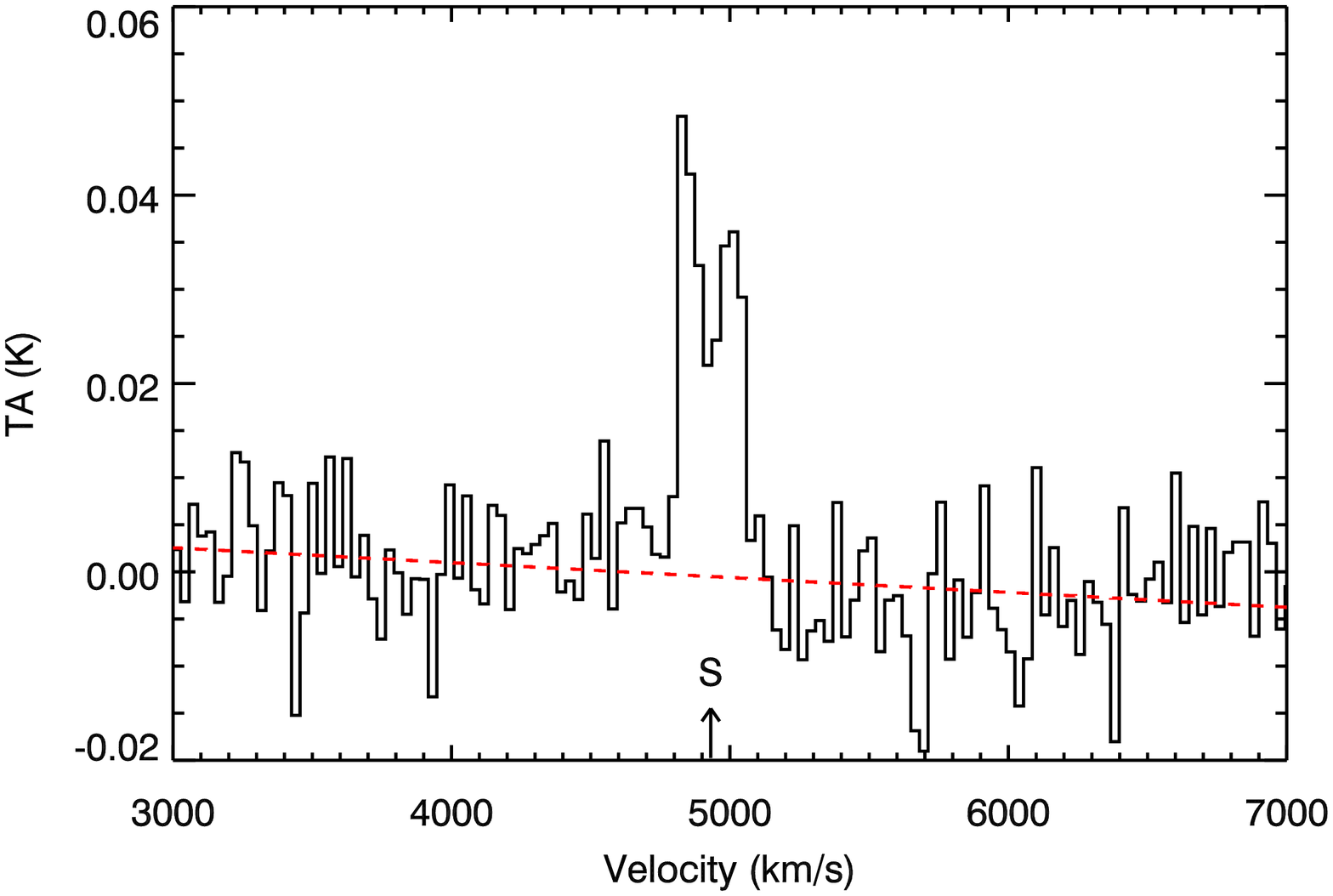}{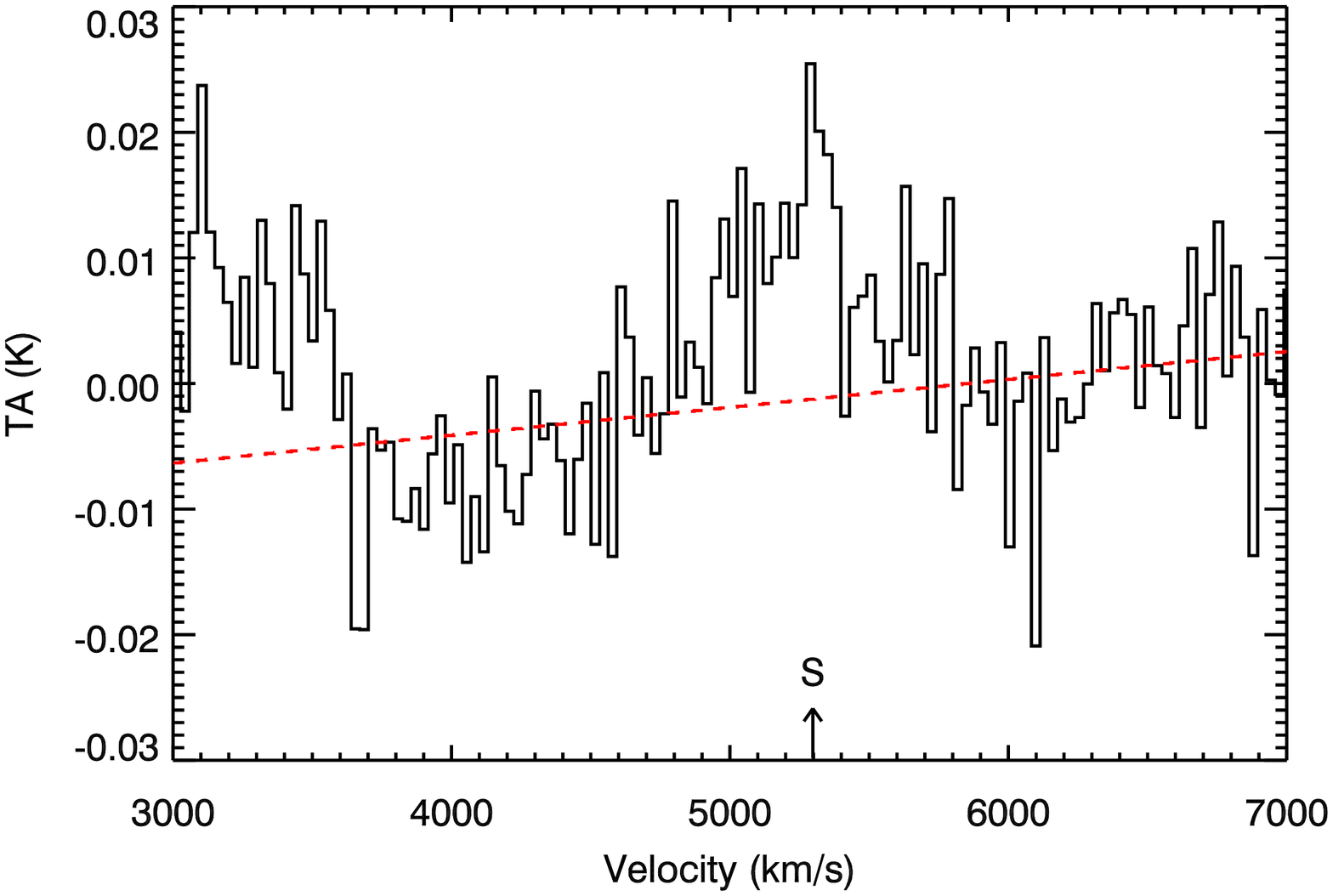}
\caption{(a)~The figure on the left shows the CO(1--0) spectrum observed from the galaxy SBS~1325+597. The emission line has a clear double
horned profile. The dashed line in red represents the baseleine fitted to the spectrum. The systemic velocity of the galaxy, $v_{sys}=4917$~kms$^{-1}$ 
(Kreckel at al. 2012) is marked with an 'S' and an arrow on the velocity axis. It lies in  
between the two emission peaks. This indicates that the molecular gas is in a rotating disk. (b)~Figure on right shows the CO(1--0) spectrum from the 
galaxy SDSSJ~143052.33+551440.0. The fitted baseline is marked with a dashed red line and the systemic velocity of the galaxy, $v_{sys}=5295$~kms$^{-1}$ 
(Kreckel at al. 2012) is marked. The emission can be clearly distinguished but is distributed 
over a range of velocities which suggests that the molecular gas distribution may not be centrally concentrated in the nucleus.}
\end{figure}

\section{Results}

\noindent
In this section we discuss our molecular gas detections (Table~1), followup H$\alpha$
observations for three of the detected galaxies (Table~3) and the implications of our results. 

\subsection{CO(1--0) detections~:~}We have detected $^{12}CO (J=1-0)$ emission from four of the five sample galaxies that we 
observed and the fluxes are listed in Table~2. The non-detection in SBS~1428+529 could be due to the short duration of the scan, 
which was limited by bad weather. The noise of the spectrum is 0.0024~K. Assuming a typical CO(1--0) bandwidth of 250~km/s we
obtain $I_{CO}~<~0.6~$K~km~s$^{-1}$, which gives a limiting molecular gas mass of $0.59\times10^{9}~M_{\odot}$ (Table~2). Of the 
four detections, SBS~1325+597 has the most striking line profile; it has a double horned structure indicating a rotating 
disk of molecular gas (Figure~1a). The velocity separation of the peaks is $\sim200~kms^{-1}$. Assuming a disk inclination of 
59.3$^{\circ}$ \citep{makarov.etal.2014} \footnote{http://leda.univ-lyon1.fr/}, the disk rotation is 116~km~s$^{-1}$. 
This is similar to the HI rotation speed in this galaxy derived from the HI 
observations of Kreckel et al. (2012). In SDSS~143052.33+551440.0 (Figures~1b) the CO line profile is fairly symmetric 
about the central systemic velocity of the galaxy but is distributed 
over a range of velocities which suggests that the molecular gas probably extends over the galaxy disk rather than concentrated
within the nucleus. In both SDSS~153821.22+331105.1 and CG~598, the molecular gas is centrally peaked (Figure~1d and 2b). SDSS~153821.22+331105.1
is barred and hence the molecular gas may have been driven into the center by the bar \citep{sakamoto.etal.1999}.

\begin{figure}
\epsscale{1.2}
\plottwo{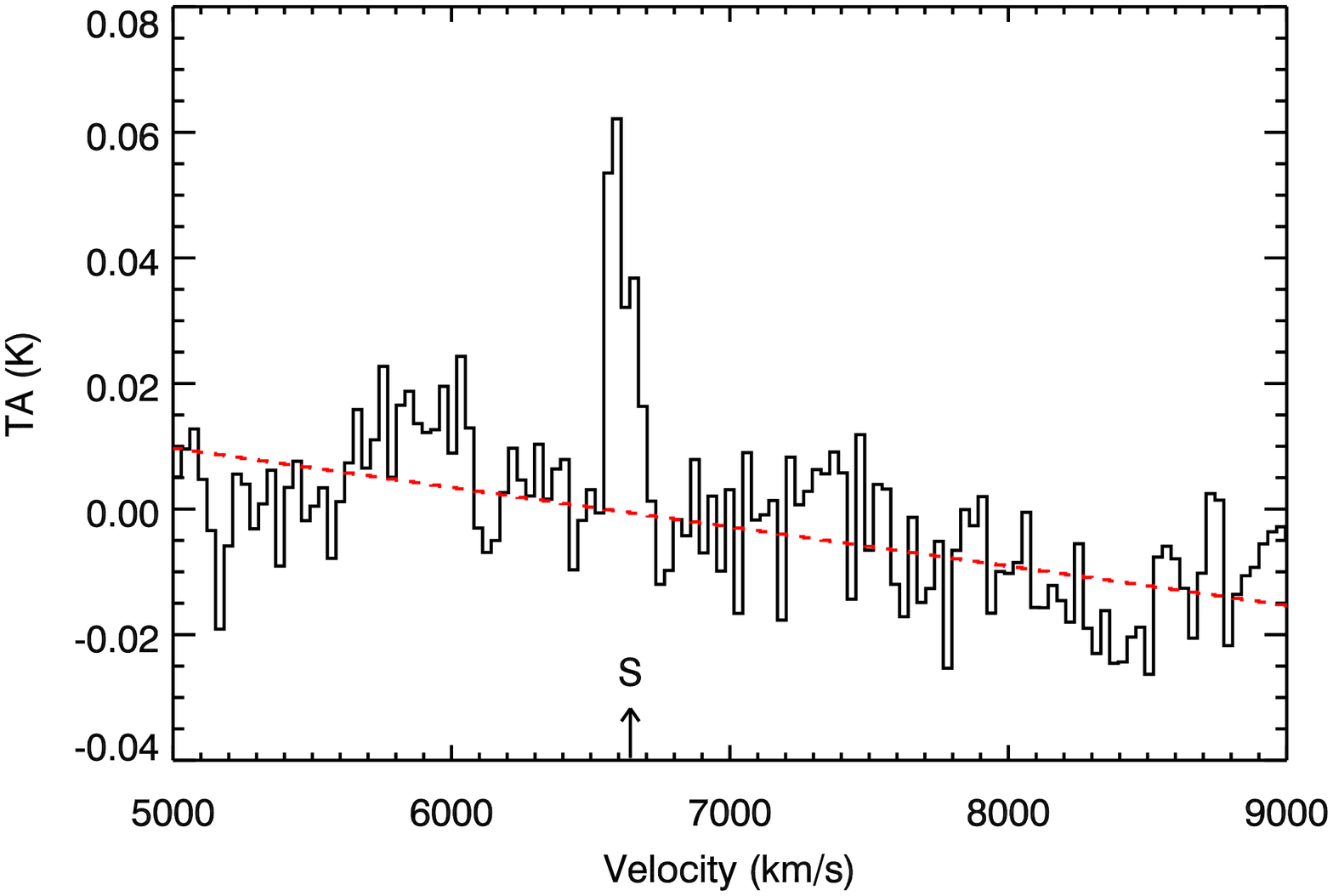}{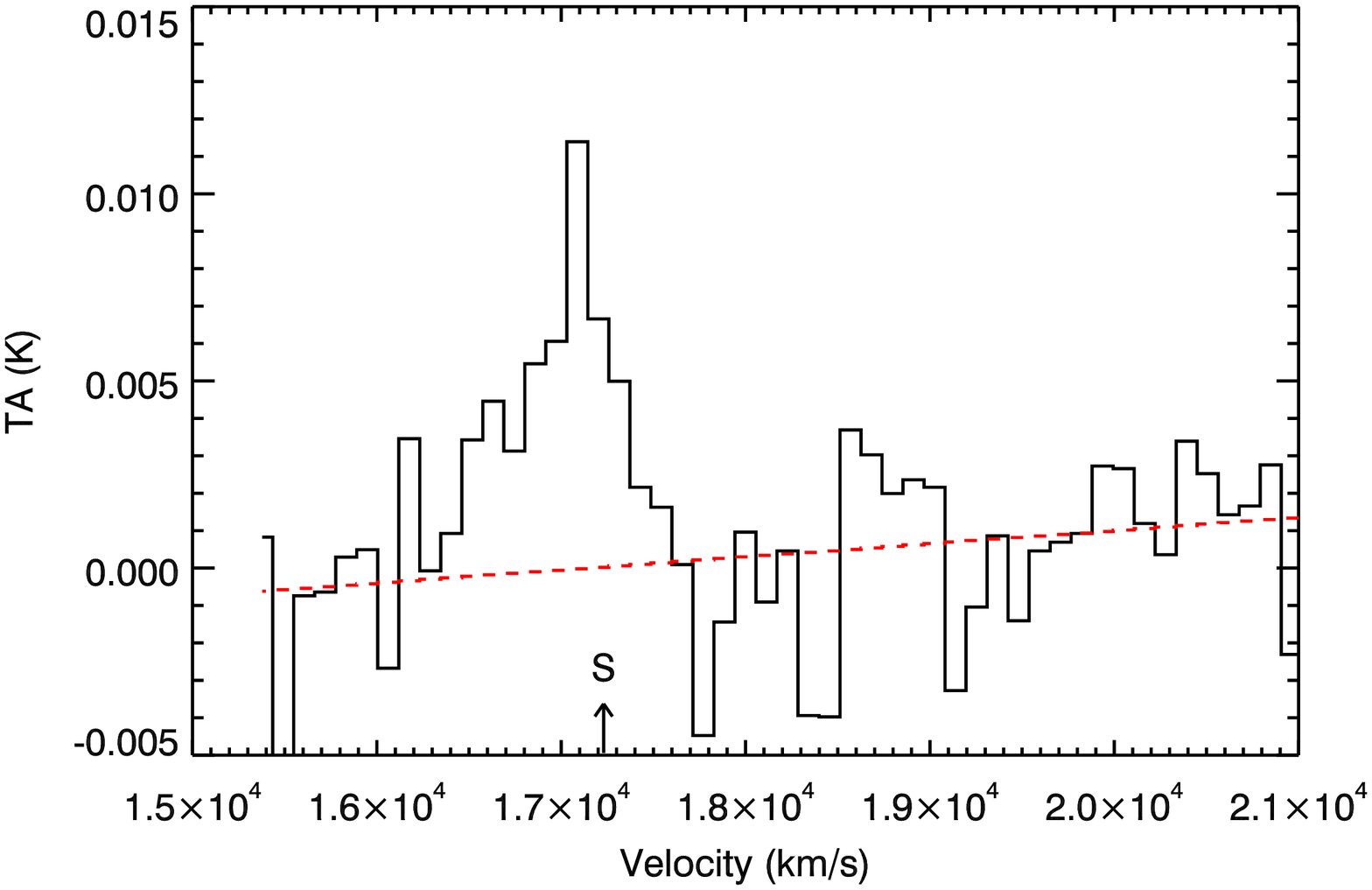}
\caption{(a)~The figure on the left shows the CO(1--0) spectrum of the galaxy SDSSJ~153821.22+331105.1. The emission is peaked about the 
systemic velocity of the galaxy, $v_{sys}=6630$~kms$^{-1}$ (Kreckel at al. 2012) which suggests that the gas is concentrated in the 
center of the galaxy. The fitted baseline is marked with a dashed red line. (b)~The figure on the right shows the CO(1--0) spectrum observed 
from CG~598. The gas distribution is also symmetric about the center of the galaxy, which is marked at $v_{sys}=17226$~kms$^{-1}$ (NED value)
with and 'S'.}
\end{figure}

\subsection{Molecular Gas Masses and surface densities~:~}The CO fluxes in K~km~s$^{-1}$ were converted to Jy~km/s using a conversion factor (Jy/K) 
of 2.4. The CO line luminosity was determined using the relation 
$L_{CO}~=~3.25\times10^{7}(S_{CO}\Delta~V/Jykms^{-1})(D_{L}/Mpc)^{2}(\nu_{res})^{-2}(1+z)^{-1}$  
and the molecular gas masses were estimated using the relation $M(H_{2})~=~[4.8~L_{CO}(K~kms^{-1}$)] (Solomon \& van den Bout 2005). 
The molecular gas masses lie in the range $(1-8)\times10^{9}~M_{\odot}$ (Table~2) which is comparable to the molecular gas masses observed 
in nearby bright galaxies that lie in denser environments \citep{helfer.etal.2003}. Using the 2MASS galaxy sizes, we estimated the suface densities
$\Sigma$ of the molecular gas distribution; it lies in the range $(3.8-30)~M_{\odot}pc^{-2}$ (Table~3), which is also similar to that observed 
in nearby bright galaxies \citep{kennicutt.1998}. Thus both the molecular gas masses and gas surface densitites are not unusually low, even though 
the galaxies are in low density environments.

\subsection{H$\alpha$ Imaging and star formation rates~:~}As mentioned earlier, due to lack of suitable filters we were able to do H$\alpha$ 
imaging of only three galaxies in our sample, SBS~1325+597, SDSS~143052.33+551440.0 and SDSS~J153821.22+331105.1 (Figures~3, 4, 5). \\
{\bf (i)}~In SBS~1325+597 the H$\alpha$ emisison is distributed over two regions on either side of the galaxy center, which indicates that it 
is only associated with the inner disk of the galaxy (Figure~3). This distribution agrees well with the CO(1--0) emission profile (Figure~1), 
which is double horned about the 
systemic velocity of the galaxy and indicates that the molecular gas is also distributed in a ring or torus about the galaxy center. In the SDSS 
optical image, it is clear that the galaxy has a red bulge but a bluish star forming disk. The H$\alpha$ luminosity gives a moderate SFR of 
0.2~M$_{\odot}$yr$^{-1}$ (Table~3). The galaxy also shows extended UV emission in its GALEX image. The emission peak is offcenter from the 
nucleus and located east of the nucleus. Thus the overall picture of this galaxy is that of an isolated void galaxy with a moderately star forming 
disk that is rich in molecular gas.\\
{\bf (ii)}~SDSS~143052.33+551440.0 is a small galaxy with a bright nucleus that appears blue in its SDSS optical image. The H$\alpha$ emisison peaks 
close to the center of the galaxy but is extended over the enitire disk (Figure~4). The H$\alpha$ luminosity gives a SFR of 0.6~M$_{\odot}$yr$^{-1}$ 
(Table~3) which is relatively strong for a galaxy of this size. Both the  H$\alpha$ flux and the CO line emission appear to peak offcenter from the nucleus.
The GALEX NUV emission is also extended over the galaxy. Thus SDSS~143052.33+551440.0 is a small, star forming, gas rich galaxy.\\
{\bf (iii)}~SDSS~153821.22+331105.1 is the only barred galaxy in our sample. It appears to be an isolated, blue, star forming galaxy with a small bulge
and faint spiral arms associated with the bar (Figure~5). The H$\alpha$ emission peaks in the nucleus and is extended over the entire bar. 
This is similar to the molecular gas distribution which is sharply peaked about the systemic velocity of the galaxy (Figure~3) indicating that the
molecular gas is concentrated in the center of the galaxy. It may have been driven into the nucleus by the bar. The bar may also have triggered the 
star formation. There is not much H$\alpha$ emission detected from the disk but the total H$\alpha$ luminosity gives a relatively high SFR of
 1.02~M$_{\odot}$yr$^{-1}$ (Table~3), probably due to the large H$\alpha$ luminosity (i.e. SFR) along the bar. There is no GALEX image for this galaxy.
Thus SDSS~153821.22+331105.1 appears to be a moderately sized barred galaxy that has strong star formation associated with the bar. It is
a good example of ongoing secular evolution in a void galaxy, where the  nuclear star formation contributes to bulge growth and the gas evolution is 
probably driven by the bar.

\subsection{Star formation efficiency and the critical mass surface density for star formation~:~}The star formation efficiecy (SFE) is 
defined as the star formation rate divided by the ratio of the molecular gas mass to the dynamical or disk rotation timescales \citep{silk.mamon.2012}. 
It is important as it indicates what fraction of the molecular gas is converted into stars during star formation in a galaxy and the gas depletion timescales 
\citep{leroy.etal.2008}. We estimated the approximate disk rotation timescales from the HI position velocity plots in Kreckel et al. (2012),
using the flat rotation velocities $v$ and HI extent $r(HI)$. We have derived the SFE in three galaxies (Table~3). For the remaining two galaxies, disk rotation 
velocities were not available in the literature and hence we could not derive their SFE's.
We used the relation SFE~=~SFR/M(H$_2$)$\times\frac{2\pi~r}{v}$ which becomes $\frac{SFR}{M(H_{2})}\frac{D^{\prime\prime}}{v(km/s)}(D_{Mpc})$
where M(H$_2$) is the molecular gas mass in solar mass units, $v$ is the depojected rotation velocities (i.e. $v_{obs}/Sin~i$), $D^{\prime\prime}$ is the 
galaxy diameter in arcseconds and $D_{Mpc}$ is the galaxy distance in Mpc. Of the three galaxies, SDSS~143052.33+551440.0 has the highest SFE, probably because it 
is much smaller than the other two galaxies and yet has comparable SFRs. It is also very blue in its SDSS composite image, which suggests a high star 
formation rate. We have also calculated the critical surface density for star formation ($\Sigma_{crit}$) \citep{kennicutt.1989}, using again the approximate 
disk rotation velocities and total gas surface densities $\Sigma(HI+H_{2}$, where $\Sigma_{crit}~=~\alpha\kappa\sigma_{gas}/3.36~G$. We assumed $\kappa=\sqrt{2}v/r$,
$\alpha=0.69$, $\sigma_{gas}=10~km/s$ where $v$ is the disk rotation velocity. For $\Sigma/\Sigma_{crit}~>~1$, the disk is unstable. Table~3 shows that
all three star forming galaxies have gas surface densitites above the critical threshold. This is in contrast to LSB galaxies that have large HI masses
but diffuse stellar disks that do not support star formation \citep{das.etal.2010}. 

\section{Comparison with previous CO detections}

There are only two other studies of molecular gas in void galaxies \citep{sage.etal.1997,beygu.etal.2013}. In the earlier study by Sage et al., 
CO(1-0) was detected from four galaxies lying within the Bootes void. Of these four galaxies, two galaxies CG~910 and CG~684 show strong CO emission 
centered around their systemic velocities. We have adjusted the distances to $H_o$~=~73~km~s$^{-1}$~Mpc$^{-1}$ and recalculated the molecular gas 
masses (Table~4). Both galaxies have molecular gas masses of the order of $10^9$ which is similar to that observed in our sample galaxies. CG~910 is 
an isolated galaxy with a relatively high disk inclination (56.2$^\circ$) and a red bulge. It has a double horned CO(1--0) emission line profile 
with peaks separated by $\sim$370~km~s$^{-1}$, which indicates that the molecular gas disk is rotating with velocity $v_{r}~=~223~km~s^{-1}$. 
The SDSS DR6 nuclear spectrum shows only weak H$\alpha$ emisison which indicates that there is only moderate star formation in this galaxy.  
The other galaxy CG~684, is closely interacting with a companion galaxy \citep{szomoru.etal.1996}. The  blue color of its SDSS
optical composite image suggests ongoing star formation \citep{cruzen.etal.2002}, that has probably been triggered by a close tidal interaction with a 
companion galaxy. The CO(1--0) 
emission line has two broad peaks separated by $\sim$710~km~s$^{-1}$, which is too large to be due to disk rotation. Instead the two peaks probably
represent gas disks in CG~684 and its companion galaxy. The strong H$\alpha$ emission line in the SDSS (DR6) spectra of CG~684 indicates that there is
ongoing nuclear star formation (Table~4). 

The second study by Beygu et al. detected  CO emission in the star forming galaxy Mrk~1477,
which is denoted by VGS~31b in their void galaxy survey \citep{kreckel.etal.2012}. The galaxy is part of a string of three galaxies that appear to lie
along an HI filament in a void. Mrk~1477 is clearly tidally interacting with the nearest galaxy VGS~31a and has a bar as well as ring in its disk.
Of the three galaxies only Mrk~1477 was studied in CO emission and the molecular gas mass is similar to our detections (Table~4). Unlike SBS1325+597 or CG~910, 
Mrk~1477 does not have a double horned CO(1--0) profile. Instead the molecular gas appears to be piled in the center of the disk. It has probably been 
driven into the galaxy center by the bar which shows strong streaming motion in H$\alpha$ \citep{beygu.etal.2013}. The nuclear SFR is significant
and has a value of 0.45~$M_{\odot}yr^{-1}$ (Table~4); the nucleus appears blue in the optical SDSS image of the galaxy. 

Thus, to date, including our detections there are seven galaxies detected in CO emission. Their molecular gas masses lie in range $10^{8}-10^{9}~M_{\odot}$, 
for g band luminosities 14.8 to 19 and SFRs 0.2 to 1.85. Also, of the seven galaxies detected in CO emission, five have blue bulges or inner disks indicating ongoing nuclear star formation. The two exceptions are CG~910 and SBS~1325+597, both of which have red bulges in their (composite) SDSS 
images. However, it is interesting to note that both CG~910 and SBS~1325+597 have double horned CO(1--0) emission profiles, which shows that their molecular gas
is distributed mainly in their disks and probably associated with disk star formation rather than nuclear star formation. The ongoing disk star formation 
in SBS~1325+597 is indicated by the faint blue color and the H$\alpha$ emission from its disk (Figure~3). A drawback of our work is 
that our sample is small, even when previous detections of molecular gas are included (Table~4) and biased towards blue galaxies. Hence, it is not representative of the entire range of void galaxies and there is also no control sample. However, when we take into account the significant amount of star formation observed in void galaxies (e.g. Kreckel et al. 2012), it is probably safe to state that molecular gas is not rare in voids. 

\begin{figure}
  \centerline{
  \includegraphics[scale=0.50,trim = 0 50 50 0,clip]{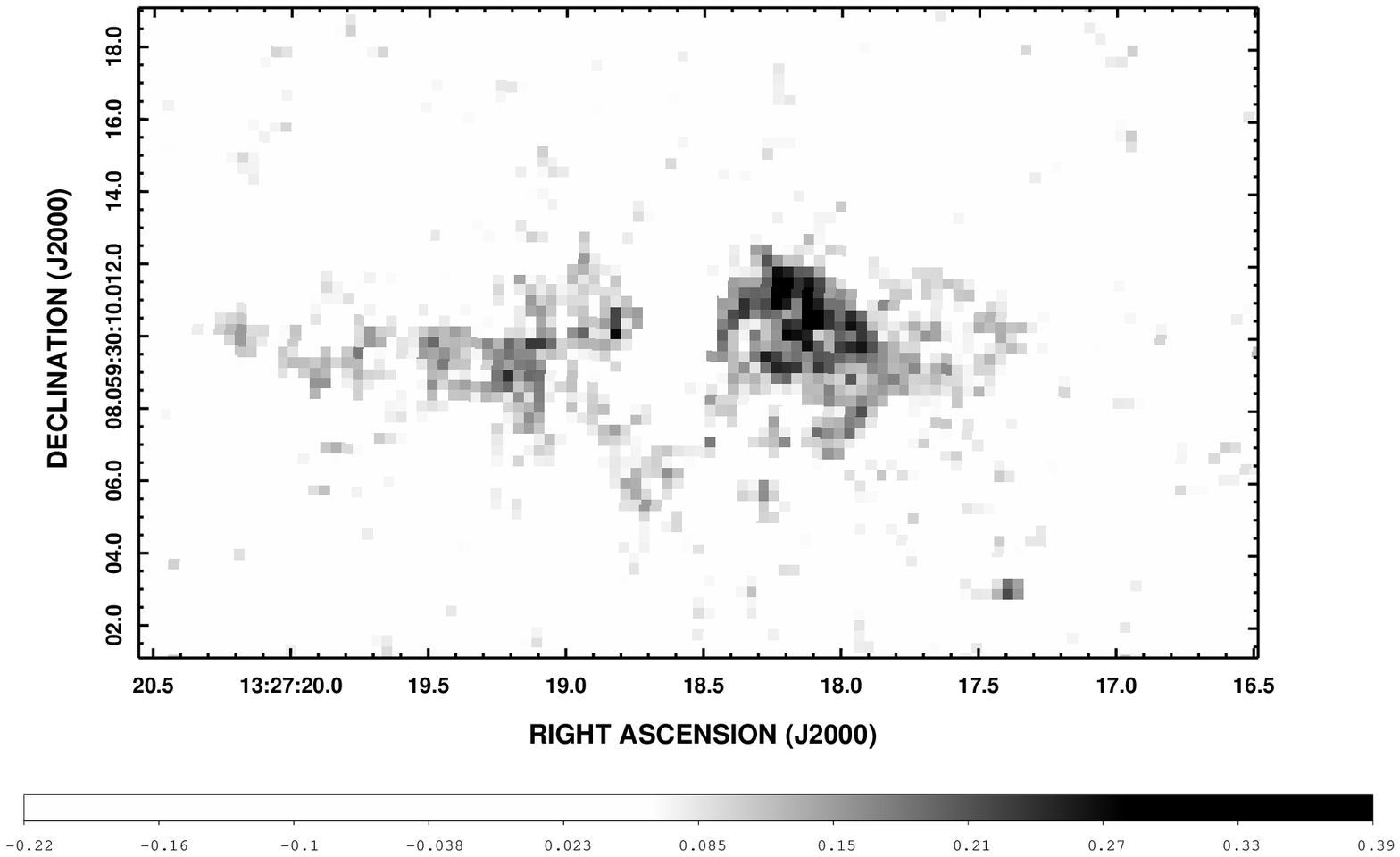}
  \quad
  \includegraphics[scale=0.50,trim = 20 50 0 0,clip]{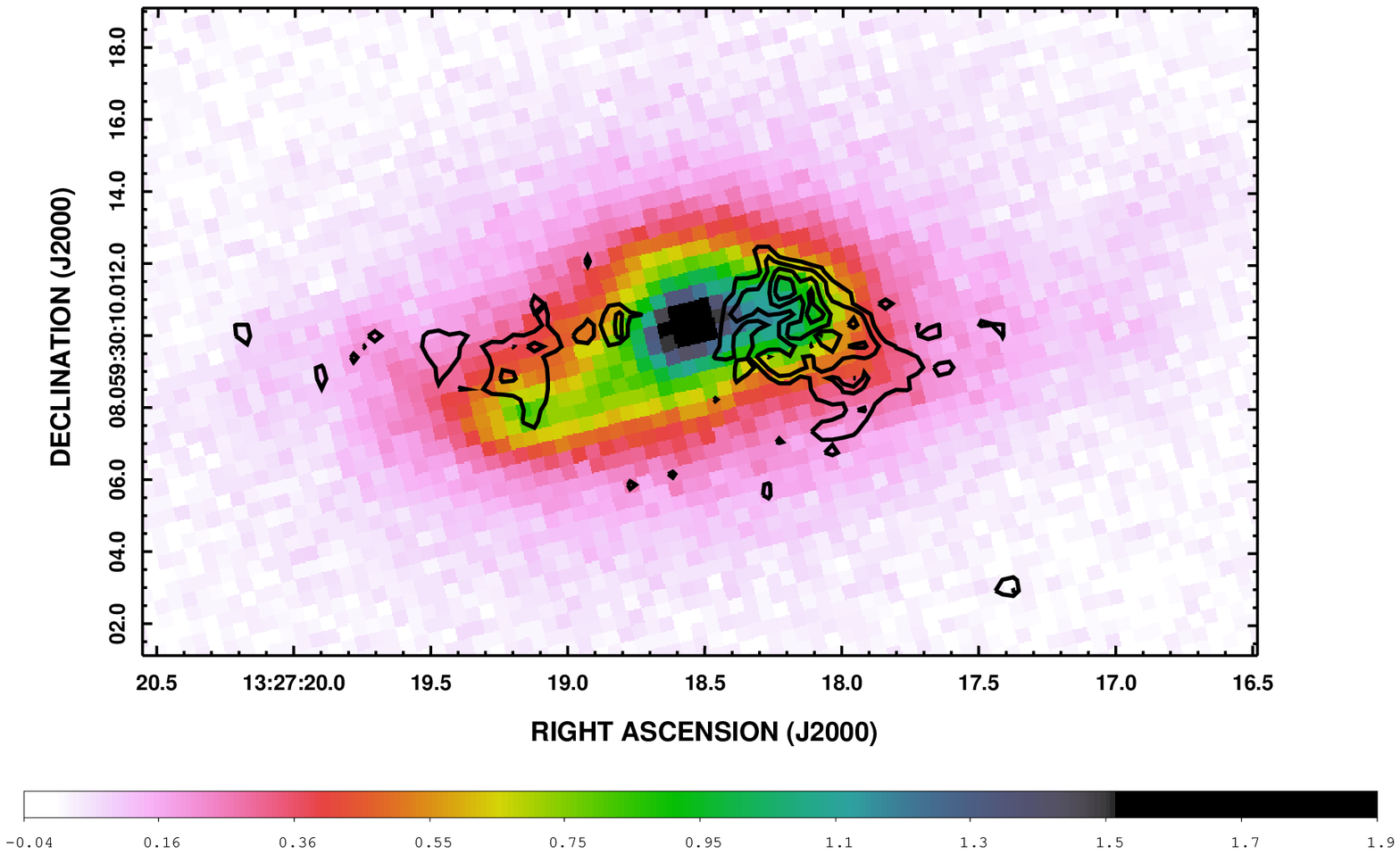}}
  \caption{(a)~The sky subtracted H$\alpha$ image of SBS~1325+597. The H$\alpha$ emission is diffuse and distributed on either side of the nucleus. There appears 
to be more star formation west of the nucleus. (b)~Contours of the H$\alpha$ emission superimposed on the SDSS g band image of the galaxy. The contour levels 
are 0, 0.12, 0.18, 0.24 and 0.30; where the units are in counts~s$^{-1}$ and the sky subtracted background has a noise level of approximately 0.03. The emission 
has a ringlike morphology and is associated with the disk. }
\end{figure}

\begin{figure}
  \centerline{
  \includegraphics[scale=0.50,trim = 0 50 50 0,clip]{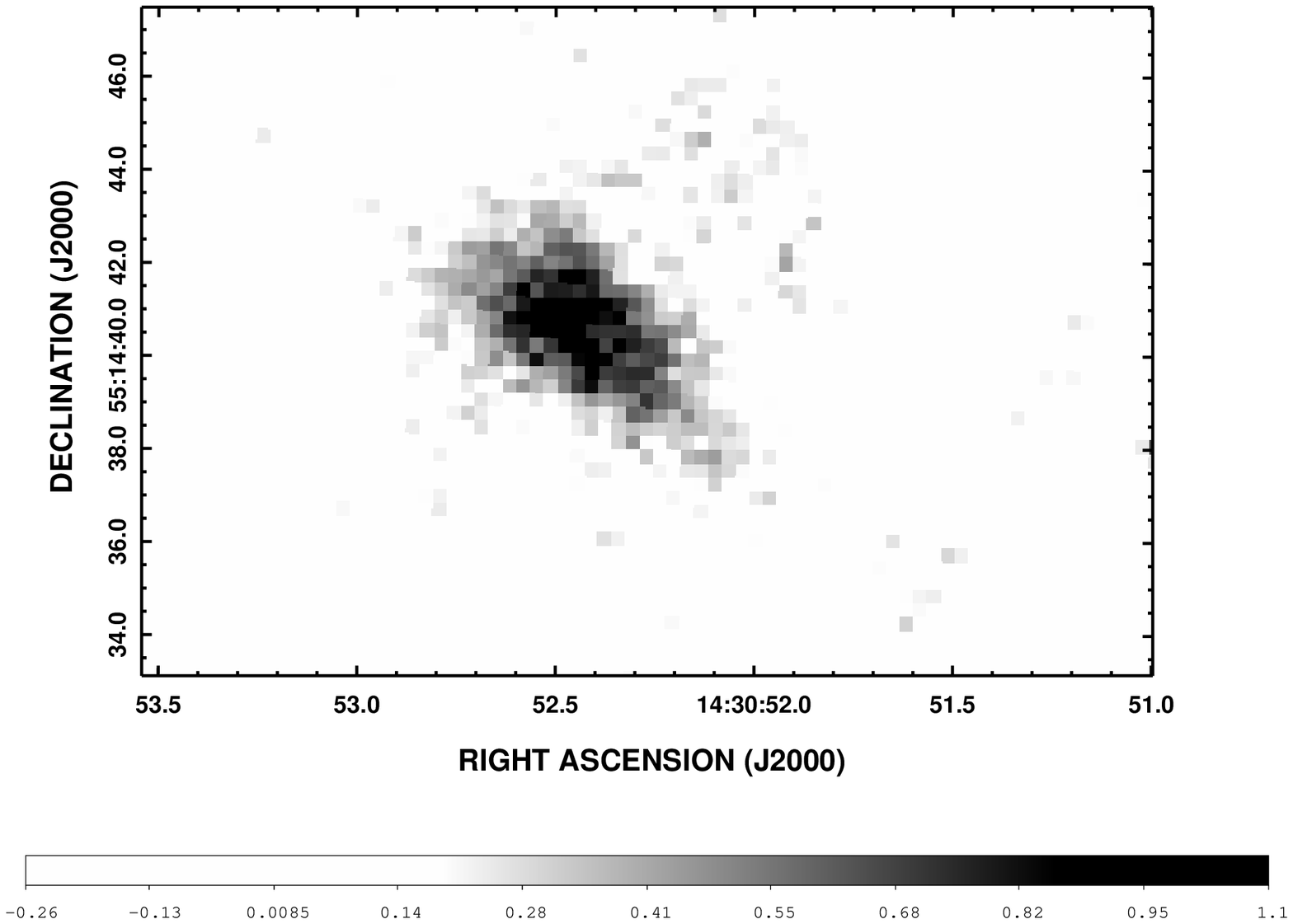}
  \quad
  \includegraphics[scale=0.50,trim = 20 50 0 0,clip]{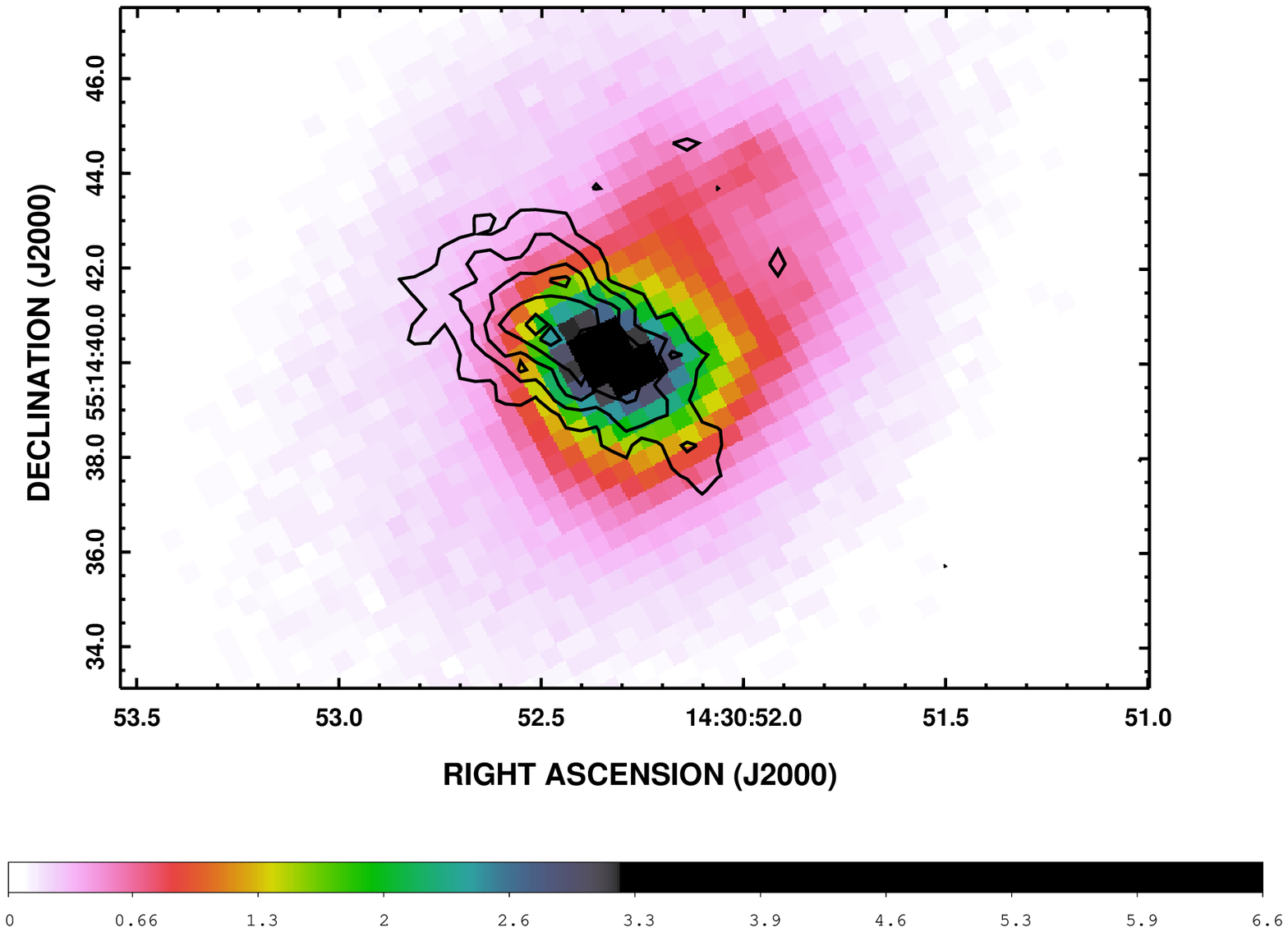}}
  \caption{(a)~The sky subtracted H$\alpha$ image of SDSS~143052.33+551440.0. The H$\alpha$ emission is distributed over the galaxy and extencec on one side. 
There is also diffuse emission lying to the northwest of the galaxy center that matches diffuse emission in the g band image as well. (b)~Contours 
of H$\alpha$ emission superimposed on the SDSS g band image of the galaxy. The contour levels are 0.27, 0.45, 0.63, 0.81 and 0.99; where the units are in 
counts~s$^{-1}$ and the sky subtracted background has a noise level of approximately 0.09. The H$\alpha$ contours are extended on one side of the nucleus.}
\end{figure}

\begin{figure}
  \centerline{
  \includegraphics[scale=0.50,trim = 0 50 50 0,clip]{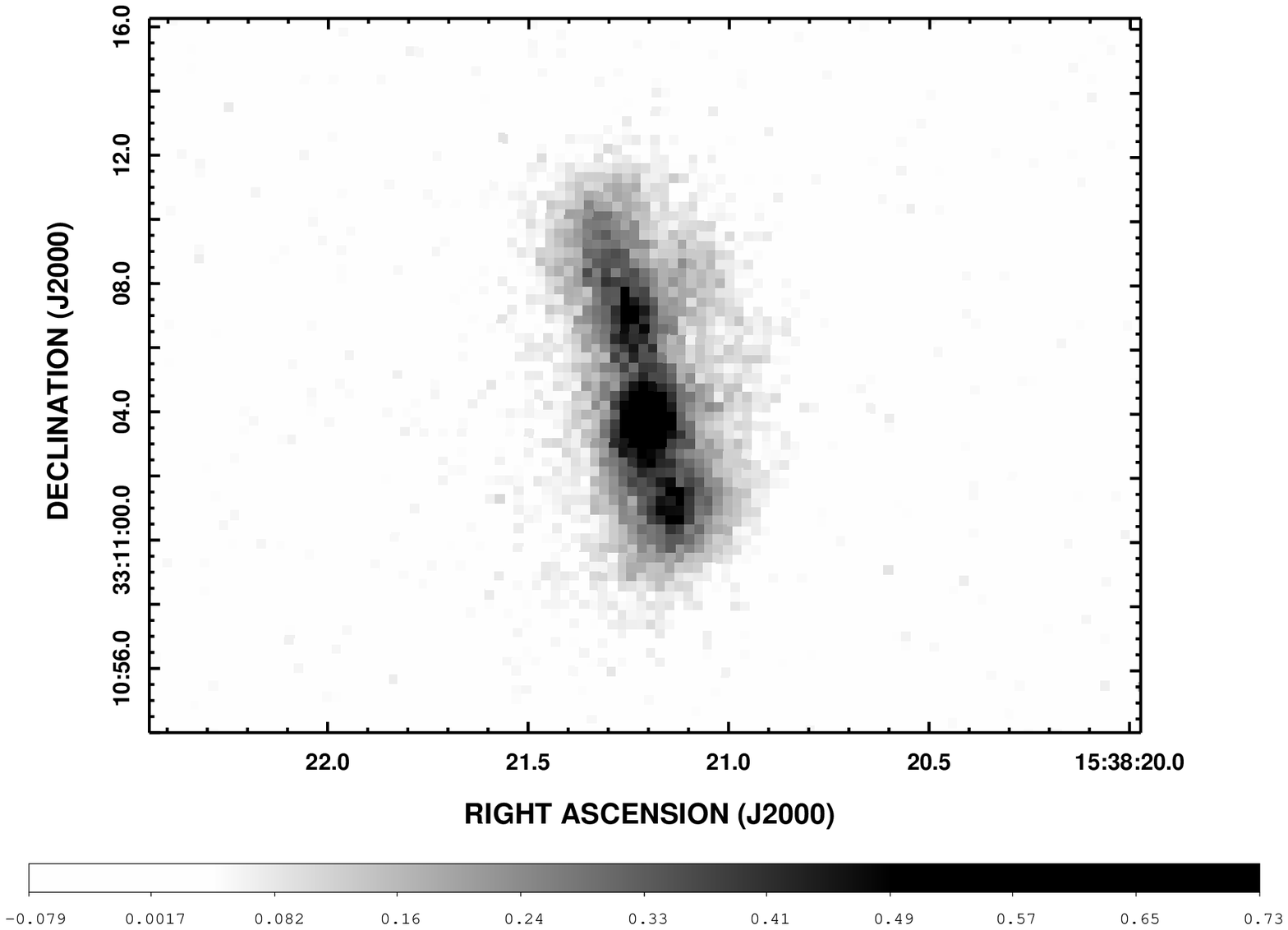}
  \quad
  \includegraphics[scale=0.50,trim = 20 50 0 0,clip]{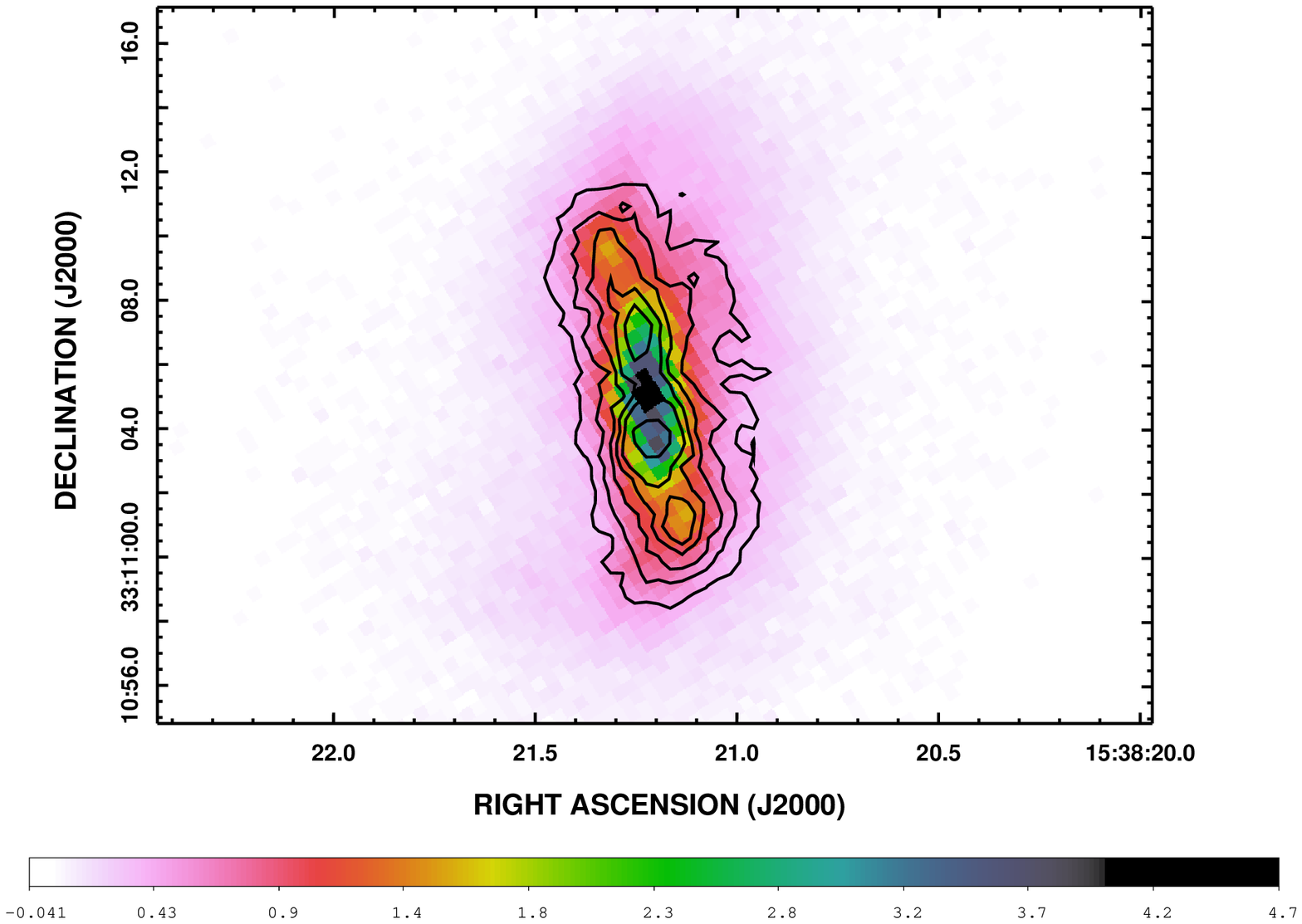}}
  \caption{(a)~The H$\alpha$ image of SDSS~153821.22+331105.1. The emission is strong and distributed along the bar in the galaxy. The flux density
is highest in the nucleus and distributed over two regions along the bar. (b)~Contours of H$\alpha$ emission superimposed on the SDSS g band image 
of the galaxy. The contour levels are 0.27, 0.45, 0.63, 0.81 and 0.99; where the units are in counts~s$^{-1}$ and the sky subtracted background has a noise 
level of approximately 0.02.  There is clearly strong star formation along the bar. The emission peaks on either side of the nucleus. }
\end{figure}

\section{Discussion}

Since molecular gas requires dust to form, its presence also indicates that 
there have been several cycles of star formation in these galaxies. In this section we address the following questions~:~(a)~what drives star formation in 
these underdense environments and (b)~are void galaxies dark matter dominated systems?\\
{\bf (i)~Galaxies and the void substructure~:~}Void galaxies form from the rare, small-scale peaks in the primordial density field and a larger volume will 
likely include more of these peaks. However, void galaxies may also form within the void substructure. In the hierarchical picture of void evolution, smaller voids merge to form larger voids \citep{sheth.weygaert.2004} causing mass to flow from the void centers to the surounding walls 
\citep{hahn.etal.2007,aragon-calvo.etal.2007,aragon-calvo.szalay.2013}. In the 
process some matter may be left behind within the voids as filaments and mini-walls tracing a void substructure 
\citep{dubinski.etal.1993,sahni.etal.1994}. In this scenerio, small groups of galaxies can form within voids at places where the filaments or 
sheets intersect each other. So even though the large scale environemnt in voids is sparse, galaxy pairs (e.g. CG~692-CG~693) or small groups 
(e.g. VGS~31a,b,c) can form within the void substructure. There is now evidence that such filamentary structures 
within voids exist \citep{alpaslan.etal.2014,beygu.etal.2013,popescu.etal.1996}. These galaxies interact or merge giving rise to star formation. 
Thus, some of the star formation that we observe in voids could be driven by the expansion and merging of the voids themselves.
Thus, it is possible that we may observe more star forming galaxies and interacting systems in the larger voids, such as the Bootes void 
\citep{szomoru.etal.1996} where many more voids could have merged to form a larger void. \\
{\bf (ii)~Slow gas accretion onto void galaxies from the IGM~:~}Slow, cold gas accretion by galaxies along filaments has been shown to be 
important for galaxies with relatively low mass halos \citep{keres.etal.2005,dekel.birnboim.2006,bouche.etal.2010}. In void galaxies, cold gas from 
the intergalactic medium (IGM) may be accreted along the filaments that form part of the void substructure. The gas accretion increases the gas
surface density as well as cools the disks. This can result in the formation of local disk instabilities which can trigger star formation. Cooler
disks are also more prone to bar instabilities, which can again trigger star formation (e.g. SDSS~153821.22+331105.1) \citep{sellwood.wilkinson.1993}.  \\ 
{\bf (iii)~Are voids dominated by LSB galaxies?~:~}$\Lambda$CDM models of structure formation predict that voids are populated mainly by dark matter dominated
galaxies, similar to the low surface brightness (LSB) galaxies that are observed at low redshifts \citep{peebles.nusser.2010}. Although the smaller voids such 
as the Local Void,  have a significant population of LSB dwarfs, the number density is not as high as predicted. Instead, surprisingly  voids contain a 
significant population of blue galaxies (Peebles 2001). The main characterstics of LSB galaxies are their extended HI gas disks and low to moderate star 
formation rates. Molecular gas is very rare in these galaxies  \citep{das.etal.2006} and their dust content is low \citep{rahman.etal.2007}. Our study 
clearly shows that molecular gas is fairly abundant in void galaxies and star formation is also present. It suggests that voids are probably not 
dominated by LSB galaxies.\\
{\bf (iv)~The dark matter content of SBS1325+597 (VGS~34)~:~}This galaxy is the only one in our sample that has a HI position velocity (PV)
plot that is extended enough to determine the flat rotation velocity in the galaxy disk.  The two horned CO emission profile 
(Figure~1a) and the PV plot in Kreckel et al. (2012) indicate a projected flat disk velocity of 100~km/s, which when  deprojected with an inclination angle of
59.3$^{\circ}$ gives a disk rotation velocity $v_c$~=~116.3~km/s. We used this velocity to obtain an approximate estimate of the 
dynamical mass. The PV plot contours extend out to an approximate radius of 
30$^{\prime\prime}$ which correponds to a lengthscale of 9.9~kpc, where 1$^{\prime\prime}$=330~pc. Thus the dynamical mass is $M_{dyn}=3.01\times10^{10}~M_{\odot}$.
To derive the stellar mass we used the (M/L)$_k$ ratios derived from closed box models of chemical evolution \citep{bell.dejong.2001}, where 
log(M/L)=a$_k$ + b$_k$(B-V). Using the (g-r) magnitudes from SDSS~DR12 and a conversion formula of (B-V)=0.98(g-r)+0.22 \citep{jester.etal.2005}, we
obtained a value (M/L)$_k$=1.07. Using the 2MASS total flux for this galaxy, we obtained a stellar mass of $M_{*}~=~1.27\times10^{10}~M_{\odot}$ for 
SBS~1325+597. As listed in Table~2, M(HI+H$_2$)=$3.84\times10^{9}~M_{\odot}$. Hence the baryonic mass is 0.55 of the total dynamical mass and so the 
galaxy is not dark matter dominated. Although this is one isolated case, the result may apply to other star forming galaxies in voids. Similar studies 
of a larger and more varied sample of void galaxies are necessary to determine the dark matter content of these galaxies. 

\section{Conclusions}

\noindent
{\bf (i)~}We searched for molecular gas in a sample of five void galaxies using the CO(1--0) emission line. We detected molecular gas in 
four of the five observed galaxies. The molecular gas masses lie between $10^{8} - 10^{9}~M_{\odot}$. \\
{\bf (ii)~}We did follow-up H$\alpha$ imaging observations of three of the detected galaxies and determined their star formation rates (SFRs)
from their H$\alpha$ fluxes. The SFR varies from 0.2 - 1~M$_{\odot}~yr^{-1}$; which is similar to that observed
in nearby star forming galaxies.\\
{\bf (iii)~}Our study and two others in the literature, indicate that although void galaxies reside in underdense 
regions, their disks contain molecular gas and may have star formation properties similar to galaxies
in denser environments. \\
{\bf (vi)~}We derived the baryonic and dark matter content of one of our sample galaxies, SBS~1325+597. We find that its baryonic content is 0.55 of
the total dynamical mass and is hence not dark matter dominated. The result may apply to other star forming galaxies in voids.



\acknowledgments
M.~Das would like to thank the anonymous referee for very useful comments that improved the paper.
T.~Saito is financially supported by a Research Fellowship from the Japan Society for the Promotion of Science for Young Scientists.
S.~Ramya kindly acknowledges the award of NSFC (Grant No. 11450110401) and President's International Fellowship Initiative (PIFI) awarded by the 
Chinese Academy of Sciences.
This work was based on observations at the Nobeyama Radio Observatory (NRO). NRO is a branch of the National Astronomical 
Observatory of Japan, National Institutes of Natural Sciences. The optical observations were done at the Indian Optical 
Observatory (IAO) at Hanle. We thank the staff of IAO, Hanle and CREST, Hosakote, that made these obervations possible.
The facilities at IAO and CREST are operated by the Indian Institute of Astrophysics, Bangalore. This research has made 
use of the NASA/IPAC Extragalactic Database (NED), which is operated by the Jet Propulsion Laboratory, California Institute 
of Technology, under contract with the National Aeronautics and Space Administration.

\noindent
Our work has also used SDSS-III data. Funding for SDSS-III has been provided by the Alfred P. Sloan Foundation, the Participating Institutions, the National Science Foundation, and the U.S. Department of Energy Office of Science. The SDSS-III web site is http://www.sdss3.org/.
SDSS-III is managed by the Astrophysical Research Consortium for the Participating Institutions of the SDSS-III Collaboration including the University of Arizona, the Brazilian Participation Group, Brookhaven National Laboratory, Carnegie Mellon University, University of Florida, the French Participation Group, the German Participation Group, Harvard University, the Instituto de Astrofisica de Canarias, the Michigan State/Notre Dame/JINA Participation Group, Johns Hopkins University, Lawrence Berkeley National Laboratory, Max Planck Institute for Astrophysics, Max Planck Institute for Extraterrestrial Physics, New Mexico State University, New York University, Ohio State University, Pennsylvania State University, University of Portsmouth, Princeton University, the Spanish Participation Group, University of Tokyo, University of Utah, Vanderbilt University, University of Virginia, University of Washington, and Yale University.



{\it Facilities:} Nobeyama Radio Observatory, Himalayan Chandra Telescope, SDSS

\bibliographystyle{apj}
\bibliography{mdas_void.1}

\clearpage

\clearpage

\begin{deluxetable}{cccccccccc}
\tabletypesize{\scriptsize}
\rotate
\tablecaption{Sample Galaxies observed in CO(1--0) emisison with NRO}
\tablewidth{0pt}
\tablehead{
\colhead{galaxy} & \colhead{RA} & \colhead{DEC} & \colhead{$D_L$} & \colhead{redshift} & 
\colhead{type} & \colhead{diameter} & \colhead{g} &  \colhead{void}\\
\colhead{name} & \colhead{(J2000)} & \colhead{(J2000)} & \colhead{(Mpc)} & \colhead{(z)} & 
\colhead{ } & \colhead{($^{\prime\prime}$)} & \colhead{magnitude} &  \colhead{name}
}
\startdata
SBS~1325+597   & 13h27m18.6s & +59d30m10s & 70.4 & 0.016 & Sm, HII & 36.0 (k$_s$) & 16.0 & Ursa Minor~I\\
SDSS~143052    & 14h30m52.3s & +55d14m40s & 76.6 & 0.018 & Extend. & 76.0 & 19.5$^d$ & Ursa Minor~I\\
SDSS~153821    & 15h38m21.2s & +33d11m05s & 97.6 & 0.022 & Sd       & 17.88 (r$^d$) & 15.3 & .... \\
CG~598         & 14h59m20.6s & +42d16m10s & 248.0 & 0.057 & HII, Sbrst & 44.7 (k$_s$) & 16.4 & Bootes\\
SBS~1428+529   & 14h30m31.2s & +52d42m26s & 191.0 & 0.044 & Sb, Sy~2   &  53.9 (k$_s$) & 15.2 & Bootes
\\
\enddata
\tablenotetext{a}{SDSS~143052 full name is SDSS~143052.33+551440.0 1 and SDSS~153821 full name is SDSS~153821.22+331105.1.}
\tablenotetext{b}{Void identifications for the first two galaxies was obtained from Kreckel et al. (2011).}
\tablenotetext{c}{Distances for all galaxies was obtained from NED, except for SDSS~143052.33+551440.0 which was obtained 
from Kreckel et al. (2012).}
\tablenotetext{d}{The optical size of the galaxy SDSS 143052.33+551440.0 1 was obtained from Kreckel et al. (2012)
and for SDSS~153821.22+331105.1 from SDSS.}
\end{deluxetable}

\begin{deluxetable}{cccccccccc}
\tabletypesize{\scriptsize}
\rotate
\tablecaption{CO emisison and derived gas masses}
\tablewidth{0pt}
\tablehead{
\colhead{galaxy}    & \colhead{t$_{obs}$} &  \colhead{S$_{CO}~\Delta~v$} & \colhead{L$_{CO}(10^{8})$} & \colhead{H$_2$ Mass} & \colhead{Surface density} 
& \colhead{HI mass} & \colhead{$\frac{M(H_{2})}{M(HI)}$}\\
\colhead{name} & \colhead{(hours)} & \colhead{(K~km~s$^{-1}$)} & \colhead{(K~km~s$^{-1}$~pc$^{2}$)} & \colhead{($10^{9}~M_{\odot}$)}
& \colhead{$\Sigma_{H_{2}}$ M$_{\odot}pc^{-2}$} & \colhead{$10^{9}~M_{\odot}$} & \colhead{}\\
}
\startdata
SBS~1325+597            & 1h~12m & 10.7~$\pm$~0.2 & (3.1~$\pm$~0.1) & 1.5~$\pm$~0.03 & 12.4~$\pm$~0.3 & 2.4~$\pm$~0.3 & 0.6 \\
SDSS~143052             & 1h~16m & 7.0~$\pm$~0.2 & (2.4~$\pm$~0.1)  & 1.1~$\pm$~0.03 & 27.5~$\pm$~0.7 & 0.50~$\pm$~0.1 & 2.3 \\
SDSS~153821             & 1h~31m & 6.4~$\pm$~0.2 & (3.5~$\pm$~0.1)  & 1.7~$\pm$~0.05 & 29.9~$\pm$~0.9 & 0.7~$\pm$~0.2 & 2.5     \\
CG~598                  & 2h~26m & 5.2~$\pm$~0.1 & (18.0~$\pm$~0.3) & 8.5~$\pm$~0.1 & 3.8~$\pm$~0.1 & 5.3~$\pm$~0.8 & 1.6 \\
SBS~1428+529            & 0h~25m & $~<~$0.6      &  $~<~$1.23       & $~<~$0.6      & .... &  2.0~$\pm$~0.4 & $~<~$0.4
\\
\enddata
\tablenotetext{a}{SDSS~143052 full name is SDSS 143052.33+551440.0 1 and SDSS~153821 full name is SDSS 153821.22+331105.1.}
\tablenotetext{b}{To derive the approximate molecular gas surface densities $\Sigma(H_{2})$ we used the galaxy diameters and distances listed in 
Table~1.}
\tablenotetext{c}{The HI masses for the galaxies SBS~1325+597 (VGS~34), SDSS~143052.33+551440.0 (VGS~44) and SDSS~153821.22+331105.1 (VGS~57) were 
obtained from Kreckel et al. (2012) and adjusted for the distances listed in Table~1. For CG~598 and SBS~1428+529 we used data from Szomoru et al.
(1996).}
\end{deluxetable}

\begin{deluxetable}{cccccccccc}
\tabletypesize{\scriptsize}
\rotate
\tablecaption{H$\alpha$ luminosities, star formation rates and star formation thresholds}
\tablewidth{0pt}
\tablehead{
\colhead{galaxy} & \colhead{H$\alpha$ Flux } & \colhead{H$\alpha$ luminosity} & \colhead{SFR} & \colhead{SFE$^b$} & \colhead{M(HI~+~H$_2$)} 
& \colhead{Disk Rotation} & \colhead{$\frac{\Sigma(HI+H_{2})}{\Sigma_{crit}}^c$} \\
\colhead{name} & \colhead{10$^{-14}$~erg~cm$^{-2}$s$^{-1}$} & \colhead{10$^{40}$~erg~s$^{-1}$} & \colhead{M$_{\odot}$yr$^{-1}$} & \colhead{}
& \colhead{$10^{9}~M_{\odot}$} & \colhead{km~s$^{-1}$} & \colhead{}\\
}
\startdata
SBS~1325+597      & 4.44     & 2.49 & 0.20     & 0.03 & 3.89  & 116.30 & 2.7 \\
SDSS~143052       & 11.65    & 7.61 & 0.60     & 0.21 & 1.60  & 145.62 & 1.5 \\
SDSS~153821       & 11.94    & 1.29 & 1.02     & 0.05 &  2.35  & 200.0  & 1.4  \\
CG~598            & ....     & ...  & 1.48$^d$ & ...  &  13.85  & ... & ... \\
SBS~1428+529      & 5.65$^e$ & 23.3 & 1.85     & .... & $<$2.61 & ... & ...   
\\
\enddata
\tablenotetext{a}{SDSS~143052 full name is SDSS 143052.33+551440.0 1 and SDSS~153821 full name is SDSS 153821.22+331105.1.}
\tablenotetext{b}{The star formation efficiency (SFE) is defined as the SFR divided by the ratio of molecular gas mass to disk rotation timescales 
(see Section~5 for details).}
\tablenotetext{c}{This is the ratio of the total gas surface density and the critical surface density for star formation 
$\Sigma(HI+H_{2})/\Sigma_{crit}~>~1$ in star forming galaxies (Kennicutt 1998). Its calculation is discussed in Section~5.}
\tablenotetext{d}{The star formation rate (SFR) is derived from the UV flux \citep{sarsyan.weedman.2009}.}
\tablenotetext{e}{The H$\alpha$ flux for SBS~1428+529 is from Weistrop et al. (1995).}
\end{deluxetable}

\begin{deluxetable}{ccccccc}
\tabletypesize{\scriptsize}
\rotate
\tablecaption{Comparison of molecular gas detection in other void galaxies}
\tablewidth{0pt}
\tablehead{
\colhead{galaxy} & \colhead{Distance} & \colhead{H$_2$ Mass}    & \colhead{HI Mass}     & \colhead{$\frac{M(H_{2})}{M(HI)}$} & \colhead{SFR} & \colhead{Reference} \\
\colhead{name}   & \colhead{Mpc}      & \colhead{10$^{9}~M_{\odot}$} & \colhead{10$^{9}~M_{\odot}$} & \colhead{}             & \colhead{M$_{\odot}$yr$^{-1}$} & \colhead{}
}
\startdata
CG~910           & 188 & 4.2  & .... & ....   & ...       &  Sage et al. (1997) \\
CG~684           & 204 & 4.7  & 9.2  & 0.51   & 1.83$^a$  &  Sage et al., Szomoru et al.(1997)\\
VGS~31b (Mrk~1477) & 69 & 1.3$^b$ & 1.5 & 0.92 & 0.49$^c$ &  Beygu et al. (2013) 
\\
\enddata
\tablenotetext{a}{The SFR was derived from the H$\alpha$ line flux $I=4.91\times10^{-14}$~ergs~cm$^{-2}$~s$^{-1}$ \citep{peimbert.etal.1992},
 using the relation relation SFR=L(H$\alpha$)/1.26$\times$10$^{42}$ (Kennicutt et al. 1998).}
\tablenotetext{b}{We have used the ${L_{CO}}^{'}$ from Beygu et al. (2013) and the relation M(H$_2$)=4.8${L_{CO}}^{'}$ to determine this value.}
\tablenotetext{c}{The SFR was derived from the SDSS DR10 H$\alpha$ line flux of $I=11.56\times10^{-14}$~ergs~cm$^{-2}$~s$^{-1}$. }
\end{deluxetable}

\end{document}